\documentclass[%
 twocolumn,
superscriptaddress,
 amsmath,amssymb,
 aps,
prb,
]{revtex4-1}

\usepackage{graphicx}
\usepackage{dcolumn}
\usepackage{bm}
\usepackage{xcolor}
\usepackage{txfonts}

\def\bk{{\mathbf{k}}}
\def\bK{{\mathbf{K}}}

\def\br{{\mathbf{r}}}

\def\bq{{\mathbf{q}}}

\def\bG{{\mathbf{G}}}

\def\beq{\begin{equation}}
\def\eeq{\end{equation}}

\def\beqa{\begin{eqnarray}}
\def\eeqa{\end{eqnarray}}

\begin{document}
	\title{Correlated insulating phases of twisted bilayer graphene at commensurate filling fractions: A Hartree-Fock study}
	
	\author{Yi Zhang}
	\affiliation{Kavli Institute for Theoretical Sciences, University of Chinese Academy of Sciences,
		Beijing, 100190, China}
	
	\author{Kun Jiang}
	\affiliation{Beijing National Laboratory for Condensed Matter Physics and Institute of Physics, Chinese Academy of Sciences, Beijing 100190, China}
	\affiliation{Department of Physics, Boston College, Chestnut Hill, MA 02467, USA}
	
	\author{Ziqiang Wang}
	\affiliation{Department of Physics, Boston College, Chestnut Hill, MA 02467, USA}
	
	\author{Fuchun Zhang}
	\affiliation{Kavli Institute for Theoretical Sciences, University of Chinese Academy of Sciences,
		Beijing, 100190, China}
	\affiliation{Chinese Academy of Sciences Center for Excellence in Topological Quantum Computation, University of Chinese Academy of Sciences, Beijing 100190, China}

\date{\today}

\begin{abstract}
Motivated by the recently observed insulating states in twisted bilayer graphene, we
study the nature of the correlated insulating phases of the twisted bilayer graphene at commensurate filling fractions.  We use the continuum model and project the Coulomb interaction onto the flat bands to study the ground states by using a Hartree-Fock approximation. In the absence of the hexagonal boron nitride substrate, the ground states are the intervalley coherence states at charge neutrality (filling $\nu$ = 0, or four electrons per moir\'e cell) and at $\nu$ = -1/4 and -1/2 (three and two electrons per cell, respectively) and the $C_2\mathcal{T}$ symmetry-broken state at $\nu$= -3/4 (one electron per cell). The hexagonal boron nitride substrate drives the ground states at all $\nu$ into $C_2\mathcal{T}$ symmetry broken-states. Our results provide good reference points for further study of the rich correlated physics in the twisted bilayer graphene.
\end{abstract}

\maketitle

	\section{Introduction}

	The discovery of flat bands and superconductivity in twisted bilayer graphene (TBG) has promoted intensive investigations of the newly layered systems coupled by weak van der Waals interaction~\cite{Cao_2018_1,Cao_2018_2,ROZHKOV20161}. In analogy to high-temperature cuprates~\cite{Lee_2006}, it was thought that superconductivity in TBG is closely related to the strongly correlated insulating phase at half filling~\cite{Cao_2018_2}. Insulating states (with and without hBN substrates alignment) have now also been observed in experiments at other fillings with integer number of electrons per moir\'e cell~\cite{Sharpe_2019,Serlin_2019, Lu_2019,Jiang_2019,Kerelsky_2019,Xie_2019,Choi_2019,Yankowitz_1059}. Many theories have been proposed to understand the superconductivity~\cite{Po_2018_1,Isobe_2018,Fan_2018,Xu_2018,Wu_2018,Lian_2019,Huang_2019,Kozii_2019,Wu_2019,Roy_2019,Tang_2019} and the correlated insulating states~\cite{Po_2018_1,Kang_2019,Zhang_2019,Isobe_2018,Nori_2019,xiao_2018,Huang_2019,Xie_2020,Dai_2019,Liu_2019,Bultinck_2019,Rademaker_2019,Wu_2020,Guinea13174,Seo_2019}.
	Theoretically, one can directly start from the tight-binding model in the original bilayer graphene lattice that captures the narrow bands, and add interaction to further study the correlating effects~\cite{Nori_2019,Rademaker_2019}. Since the unit cell in such models contains tens of thousands of atoms, it can raise numerical difficulties when dealing with long-ranged Coulomb interaction. In order to overcome this difficulty, two kinds of methods are applied to construct the TBG effective models.
	One is to use the continuum model~\cite{BM_2011,Lopes_2012} by scattering between Dirac points that belong to different layers. And the other is to construct tight-binding models from local Wannier orbitals~\cite{Yuan_2018,Kang_2018,Koshino_2019,Po_2019,Carr_2019}.

	In this work, we use the continuum model~\cite{BM_2011,Lopes_2012} to carry out a self-consistent mean-field study, which is equivalent to the Hartree-Fock approximation, for TBG commensurate fillings fractions corresponding to integer numbers of electrons per moir\'e cell. Due to particle-hole symmetry, we only need to consider fillings $\nu=-3/4$, $-1/2$, $-1/4$ and $0$ for one, two, three and four electrons occupying the 8 moir\'e flat bands, respectively. The ground states at positive fillings can be obtained by particle-hole symmetry.  The flat bands are characterized by spin s, valley $\tau$ and band index m.  In addition to the spin degrees of freedom, there are two sets of flat bands, which can be denoted as $\tau=\pm$ , resulted from the interlayer scattering of the states around the Dirac points at the graphene valley $K^{\pm}$, as shown in Fig.1(b).

	\begin{figure}[h!]
		\includegraphics[width=\columnwidth,clip=true]{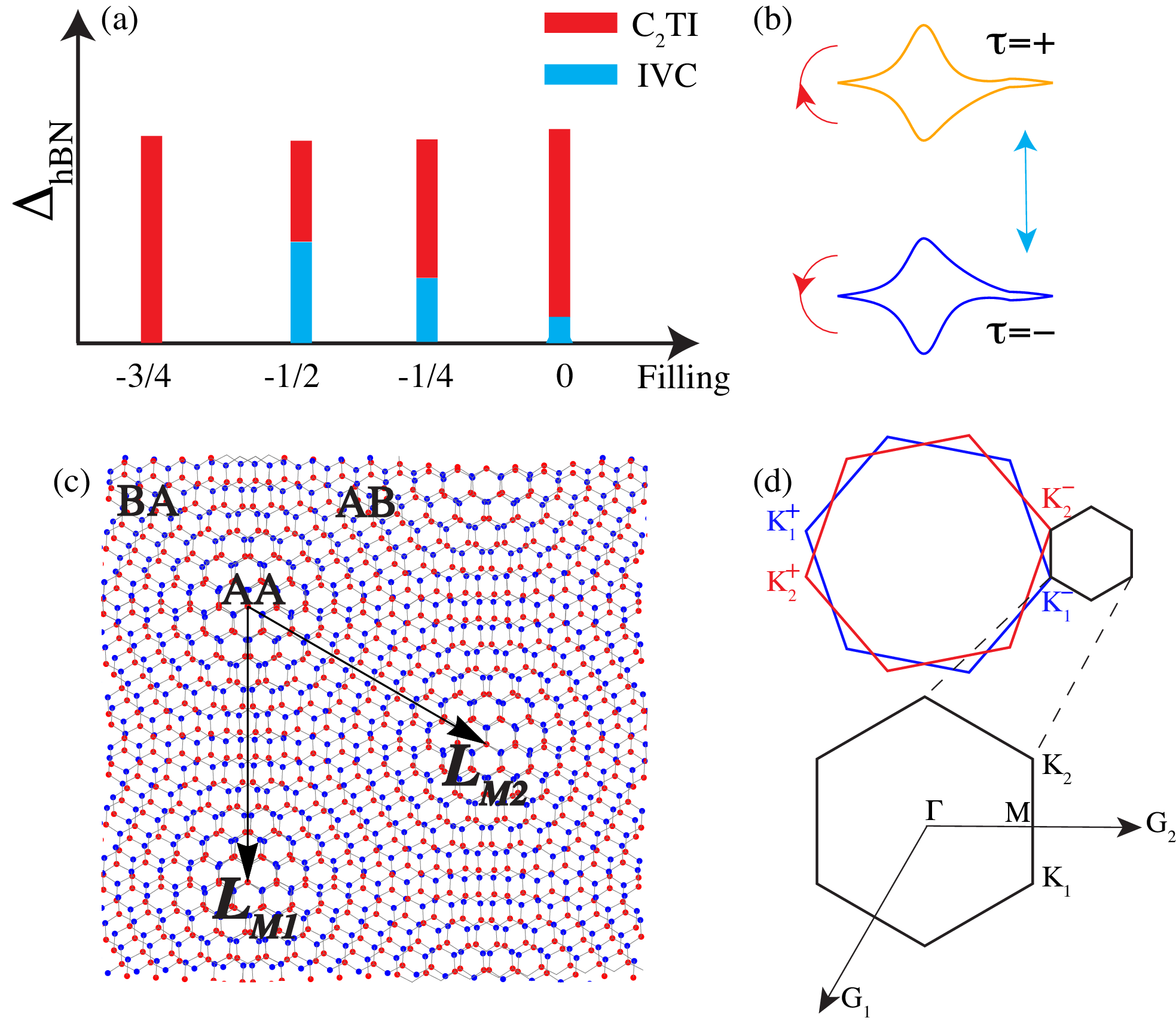}
		\caption{(a) Schematic phase diagram of TBG in the plane spanned by commensurate filling fraction and hBN potential $\Delta_{hBN}$. The red color stands for $C_2\mathcal{T}$ breaking insulator ($C_2\mathcal{T}$I) while the cyan for the intervalley coherent (IVC) state. (b) TBG contains two sets of flat bands from valley $\tau=+$ and $\tau=-$. IVC state is favored by the intervalley scattering (cyan arrow). The $C_2\mathcal{T}$I is favored by the interband scattering in each valley (red arrow). (c) Atomic structure of the TBG with twisted angle $\theta$=6.01$^{\circ}$, where the red and blue dots represent the
		A, B sublattice. The AA region corresponds to the region where the two red dots from the two layers lie on top of each other while the AB and BA correspond to the region where the red and blue dots overlap with each other.
		(d) Schematic plot of the Brillouin zone folding in TBG, where the small hexagon represents the moir\'e
		Brillouin zone reciprocal to the moir\'e superlattice structure in (c). $\bG_1$ and $\bG_2$ are the reciprocal lattice vectors of the moir\'e Brillouin zone.}
		\label{fig:fig1}
	\end{figure}
	
	We project the Coulomb interaction between electrons into the flat band basis. This process is formally identical to the projection of the Coulomb interaction to the lowest Landau level in the fractional quantum Hall effect~\cite{Haldane_1983}. Then, we can apply a Hartree-Fock mean-field theory by decoupling the projected interactions into momentum dependent order parameters.
	The main results are summarized in Fig.1(a). There are mainly two kinds of states, intervalley coherence states (IVC) and the $C_2\mathcal{T}$ breaking states ($C_2\mathcal{T}$I). The intra-valley interband scattering favors $C_2\mathcal{T}$I (red arrow in Fig.1(b)) while the intervalley scattering favors IVC states (cyan arrow in Fig.1(b)).
	If the substrate potential $\Delta_{hBN}$=0, IVC states are found to be the ground state for $\nu=$ $-1/2$, $-1/4$ and $0$, while $C_2\mathcal{T}$I states are found to be the ground states for $\nu=$ -3/4.
	By increasing $\Delta_{hBN}$, $C_2\mathcal{T}$I states become more stable comparing to IVC states.
	Decreasing the bandwidth of the moir\'e bands reduces the energy difference between the IVC state and $C_2\mathcal{T}$I state, but does not affect the ground state until the moir\'e bands become completely flat where the two states are degenerate in energy.
	
	Recently, the continuum model based Hartree-Fock approximation has been carried out by several groups. Xie et al.~\cite{Xie_2020} and Liu et al.~\cite{Dai_2019} have adopted the Hartree-Fock calculations that include flat bands as well as many remote bands to address the correlation and topological properties of TBG.
    Specificly, the latter introduces the momentum independent order parameters but studies more complete phase diagram at $\pm1/2$ and $\pm3/4$ fillings. Moreover Liu et al.~\cite{Liu_2019}, Bultinck et al.~\cite{Bultinck_2019} have also applied a similar approximation by projecting the interactions to the lower energy bands to study the insulating phase at the charge neutral point.
	
	This work is organized as follows. We start from the discussion of the continuum model and the Hartree-Fock mean-field formulation in Section~\ref{sec:formulation}. In Section~\ref{sec:results}, we discuss the main results at charge neutrality $\nu$=0, half-filling $\nu$=1/2 and $\nu$=3/4 filling. Then, we will briefly mention the results at $\nu$=1/4 filling and finalize the discussion in Section~\ref{sec:conclusion}.

\section{Formulation}
\label{sec:formulation}

\subsection{moir\'e lattice structrue}
The moir\'e pattern is formed by twisting the top and bottom layers of an
aligned bilayer graphene by angles $\theta/2$ and $-\theta/2$. As shown in the panel (c) in Fig.~\ref{fig:fig1}, the formed
periodic superlattice structure can be viewed as a triangular lattice of
the AA region, where the atoms from the same sublattice of the two layers
lie on top of each other with the lattice vectors
$\mathbf{L_{M1}}=(0,-1)L_M$, $\mathbf{L_{M2}}=(\frac{\sqrt{3}}{2},-\frac{1}{2})L_M$ and
$L_M=a_0/(2\sin(\theta/2))$, where $a_0$=0.246 nm, is the lattice
constant of the monolayer graphene.
The corresponding reciprocal lattice vectors of the moir\'e lattice are
$\bG_1=(-\frac{2\pi}{\sqrt{3L_M}},-\frac{2\pi}{L_M})$ and
$\bG_2=(-\frac{4\pi}{\sqrt{3L_M}},0)$.
After the twisting, the momentum of the two Dirac points of the two
layers becomes $\bK_1^{\tau}$ and $\bK_2^{\tau}$ and they are equivalent
to the $-\tau\bK_1$ and $-\tau\bK_2$ in the moir\'e Brillouin zone as
shown in Fig.~\ref{fig:fig1}(d).

\subsection{Continuum model}
\label{subsec:Continuum}

The low energy physics of TBG can be described by the continuum model introduced by
Bistritzer and MacDonald~\cite{BM_2011,Lopes_2012}. Here we formulate the continuum model as described by the Hamiltonian,
\begin{equation}
    H^{\tau}_{BM}(\hat{\mathbf{k}})=\left(\begin{array}{cc}
-\hbar v_F (\hat{\bk}-\bK_1^{\tau}) \cdot \pmb{\sigma}^{\tau}  &  U  \\
U^+  & -\hbar v_F (\hat{\bk}-\bK_2^{\tau}) \cdot \pmb{\sigma}^{\tau}
\end{array}\right)
\label{eq:model}
\end{equation}
where $\tau=\pm$ is the valley index,
$\pmb{\sigma}^{\tau}=(\tau \sigma_x,\sigma_y)$ which is the Pauli matrices
defined in the A,B sublattice space.
$\bK^{\tau}$ correspond to the two inequivalent Dirac points of the unrotated monolayer graphene, while
$\bK^{\tau}_1$ and $\bK^{\tau}_2$ are the corresponding Dirac points of the bottom and top layers that are twisted by angles $\mp\frac{\theta}{2}$, and
$\hat{\bk}=-i\partial_{\br}$.
The interlayer tunneling between the the Dirac states in the two layers is described by the matrix
\begin{equation}
\begin{split}
    U=&
    \left(\begin{array}{cc}
    u_0  & u_1 \\
    u_1 & u_0
    \end{array}\right)
+   \left(\begin{array}{cc}
    u_0  & u_1 e^{-\tau \frac{2\pi}{3}} \\
    u_1 e^{\tau \frac{2\pi}{3}} & u_0
    \end{array}\right)e^{-i \tau \bG_1 \cdot \br}\\
&+   \left(\begin{array}{cc}
    u_0  & u_1 e^{\tau \frac{2\pi}{3}} \\
    u_1 e^{-\tau \frac{2\pi}{3}} & u_0
    \end{array}\right)e^{-i \tau(\bG_1+\bG_2) \cdot \br}
\end{split}
\end{equation}
where $u_0$ and $u_1$ are the intra-sublattice and inter-sublattice interlayer tunneling amplitudes.
This continuum Hamiltonian is spin independent, so it has two SU(2) symmetries at
each valley. Since it does not have terms that couple the two valleys, it also has a
$U_v(1)$ for valley charge conservation in addition to the $U_c(1)$ for total charge conservation symmetry.
Moreover, in the diagonal blocks of Eq.~\ref{eq:model}, we have neglected the rotation of the momentum $\hat{\bk}-\bK_{1,2}^{\tau}$ about the z axis for $\pm\theta/2$, due to its tiny effect for small twisted angle $\theta$, which leads to an additional particle-hole symmetry~\cite{Bultinck_2019,Dai_2019}. We have checked that the inclusion of such symmetry breaking term will not affect the results of the paper.
The eigenstate of $H_{BM}^{\tau}$ can be written in the Bloch wavefunction form
\begin{equation}
 \psi_{m,\tau,\bk}^X(\br)=\sum_{\bG}u_{m,\tau;\bG,X }(\bk) e^{i(\bk+\bG)\cdot \br}
 \label{eq:bloch}
\end{equation}
where $X=\{A_1, B_1, A_2, B_2\}$ is the layer and sublattice index with the eigen-energy $\epsilon_{m\bk\tau}$.
Here, m and $\tau$ are the band and valley indices and we omit the spin
index s here since the Hamiltonian is spin independent.
Throughout the paper, we fix the parameters as $\hbar v_F/a_0$=2.365 eV,
the twist angle is fixed at $\theta$=1.086$^{\circ}$.
The inter-sublattice interlayer coupling is chosen as $u_1$=0.11 eV, so that the moir\'e bands are completely flat when
the intra-sublattice interlayer coupling $u_0$ vanishes~\cite{Tar_2019}.
In order to see the effect of moir\'e bandwidth, we tune the $u_0$ from the
realistic value~\cite{Carr_2019,Nam_2017} of 0.08 eV which corresponds to a bandwidth of 1.8 meV to
the value of 0.01 eV resulting in a bandwidth of 0.03 meV.

\subsection{Mean field theory}
In order to study the effects of the Coulomb interaction, we apply our mean-field theory in the momentum space to decouple the interaction term.
We solve a set of self-consistent equations to search for possible symmetry breaking
states due to the Coulomb interaction.
The mean-field calculation is performed in momentum space, which allows us to avoid the real space Wannier obstruction
~\cite{Po_2018_1,Po_2018_2,Po_2019,Ahn_2019}.
Moreover, in our calculation, we project
the Coulomb interaction onto the two flat bands.
More details of the projection to the flat bands are provided in
appendix~\ref{sec:app_proj}.

After projecting onto the two flat bands, the total Hamiltonian becomes
\begin{equation}
\begin{split}
    H&=\sum_{m\bk s\tau}(\epsilon_{m\bk\tau}-\mu)d_{m\bk s\tau}^{+}d_{m\bk s\tau}
    \\
    & + \frac{1}{2S} \sum_{\{m_i\}}\sum_{ s s'\tau\tau'}\sum_{\bk_1\bk_2\bq}V_{m_1,m_2,m_3,m_4}^{\tau\tau'\tau'\tau}(\bk_1,\bk_2,\bq)
    \\
    & d_{m_1\bk_1-\bq  s\tau}^+d_{m_2\bk_2+\bq  s'\tau'}^+d_{m_3\bk_2  s'\tau'}d_{m_4\bk_1  s\tau}
\end{split}
\label{eq:Hproj}
\end{equation}
where S is the total area, $d^{+}_{m\bk  s\tau}$ and $d_{m\bk  s\tau}$ are the creation and annihilation operators in the projected flat band basis.
Here, we neglect the intervalley interaction due to the Coulomb scattering
between the two valleys $V(\bK^{+}-\bK^{-})$ whose ratio to the intravalley
interaction can be estimated by $a/L_M$ which is negligible for small twist angle.
The projected interaction is derived from the Coulomb interaction $V(\bq)$ as
\begin{equation}
\begin{split}
 &V_{m_1,m_2,m_3,m_4}^{\tau\tau'\tau'\tau}(\bk_1,\bk_2,\bq)=\sum_{\bG} V(\bq+\bG)
 \\
 &
 \lambda_{m_1 m_4; \tau}(\bk_1-\bq,\bk_1+\bG) \lambda^{*}_{m_3 m_2; \tau^{'}}(\bk_2,\bk_2+\bq+\bG)
\end{split}
\label{eq:hproj}
\end{equation}
where $\lambda_{m_1 m_2; \tau}(\bk_1,\bk_2+\bG)$ is the form factor written in terms of the Bloch wavefunction Eq.~\ref{eq:bloch} as
\begin{equation}
 \lambda_{m_1 m_2; \tau}(\bk_1,\bk_2+\bG)=\sum_{\bG^{'},X}u^{*}_{m_1,\tau;\bG^{'},X}(\bk_1)u_{m_2,\tau;\bG+\bG^{'},X}(\bk_2)
 \end{equation}
We take $V(\bq)$ as the single-gate-screened Coulomb potential~\cite{Liu_2019,Bultinck_2019}
\begin{equation}
 V(\bq)=\frac{e^2}{2\varepsilon\varepsilon_0 q}(1-e^{-2q d_s})
 \label{eq:Vg}
\end{equation}
In this paper, we take the dielectric constant $\varepsilon=7$ and the gate distance
$d_s=40$ nm. We have also checked the results using other sets of parameters and
it will not change the ground state of the system.

Next, we perform a standard self-consistent mean-field calculation on Eq.~\ref{eq:Hproj}.
Since we only consider the states that preserve the translation symmetry at the
scale of moir\'e unit cell, the mean-field order parameters are then defined by the density matrix $\rho(\bk)$ whose elements are
\begin{equation}
    \rho(\bk)_{m_1, s\tau;m_2, s'\tau'}= \left<d_{m_1\bk s\tau}^{+}d_{m_2\bk s'\tau'}\right>
\label{eq:op}
\end{equation}
Since this density matrix is finite even for the nonsymmetry breaking states in the charge channel, which will be double-counted when coupled to the density
operators in the mean-field Hamiltonian, we need to remove these double counting
terms in the calculation. Different approaches have been used to address this issue~\cite{Xie_2020,Liu_2019,Bultinck_2019}. Here we take care of this by subtracting the density matrix of the
nonsymmetry breaking states $\rho_0(\bk)$ at the filling studied from $\rho(\bk)$,
so that there is always a trivial solution from the mean-field Hamiltonian which corresponds to the fully symmetric state.
This means the density matrix that couples to the
density operator in the mean-field Hamiltonian is effectively
$\rho(\bk)-\rho_0(\bk)$.
More details of the mean-field approximation are provided in appendix~\ref{sec:app_hf}.
At charge neutrality, our approach is equivalent to that used in Ref.~\onlinecite{Liu_2019}.

\section{Results}
\label{sec:results}
We have applied the self-consistent mean-field calculation to different integer fillings, and find several symmetry breaking gapped states induced by the interaction.

\subsection{$\nu=0$ case}
At the charge neutral point, there are four electrons
occupying the eight moir\'e bands. The self-consistent mean-field calculation
leads to several symmetry breaking insulating states
which can be classified into three groups
according to the specific symmetry that is broken in these states.
The first group is the flavor-polarized states including spin-polarized (SP),
valley-polarized (VP) and spin-valley locked (SV) states
which either break $SU(2)$ symmetry or spinless time reversal symmetry
$\mathcal{T}$ or both.
These states are generalized ferromagnetic insulating states (FMI).
If we let $s$, $\tau$ and $\gamma$ be the Pauli matrices in the spin, valley and
band basis the order parameter of the FM states can be written as
$s_z\tau_0\gamma_0$, $s_0\tau_z\gamma_0$ or $s_z\tau_z\gamma_0$
and these states are degenerate in energy.
The second group includes the insulating states that break $C_2\mathcal{T}$ symmetry and is
labeled by $C_2\mathcal{T}$I, where the Dirac points are completely gapped out
due to the $C_2\mathcal{T}$ breaking.
Their order parameter is dominated by the term
$s_{0,z}\tau_{0,z}\gamma_{x}$. They are also degenerate in energy.
The last group is the intervalley coherent (IVC) state, which mixes the
states from the two opposite valleys so that the $U_v(1)$ valley charge convervation
is broken. The order parameter associated with this state can be written as
$s_{0,z}\tau_{x,y}\gamma_{y}$.

\begin{figure}[h]
\includegraphics[width=\columnwidth,clip=true]{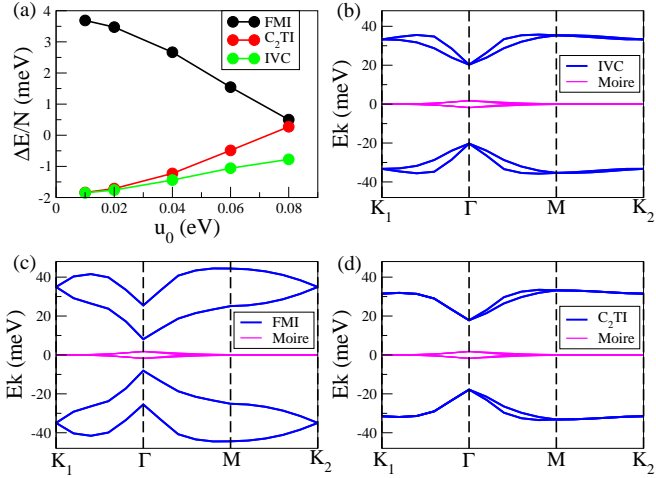}
  \caption{(a) Total energy per particle for ferromagnetic insulator (FMI), $C_2\mathcal{T}$ breaking insulator ($C_2\mathcal{T}$I) and intervalley coherent (IVC) states at the
  charge neutrality for various values of the intra-sublattice interlayer coupling
  $u_0$, where we choose $\Delta E$ to be relative to the average energy of these three states to clarify the small energy difference. (b)(c)(d) Typical energy dispersion of states of these three groups of states, where
  $u_0$=0.08 eV and the chemical potential is moved to the center of the gap for convenience.
  The non-interaction bands calculated from the $H_{BM}$ are also plotted for comparison.}
  \label{fig:energy_cn}
\end{figure}

In order to determine which state is the ground state, we calculate the total
energy per particle for each mean-field solution. As shown in Fig.~\ref{fig:energy_cn}, the IVC state
is always the ground state. For realistic parameters, where the intra-sublattice
interlayer coupling $u_0$=0.08 eV, the IVC state is about 1 meV lower than the
$C_2\mathcal{T}$I and FMI states. As $u_0$ decreases from 0.08 eV to 0.01 eV,
where the bandwidth of the non-interaction moir\'e bands
becomes flatter and flatter, the energy of $C_2\mathcal{T}$I
becomes closer and closer in energy to the IVC state. These results are consistent with
those of Ref.~\onlinecite{Liu_2019,Bultinck_2019}.
The IVC has an energy gap of about 40 meV as shown in the panel (b) in
Fig.~\ref{fig:energy_cn}. Panel (c) shows one of the (VP) FMI states  where
all the four bands have double spin degeneracy. Panel (d) shows one of the
$C_2\mathcal{T}$I states with order $s_0\tau_0\gamma_x$, which has an energy
gap of about 35.5 meV.

We have also studied the effect of sublattice potential, which is associated
with the alignment of the hBN substrate with one of the graphene layers.
This staggered potential can be formulated by an extra term $\Delta_{hBN}\sigma_z$
adding to Eq.~\ref{eq:model}.
We find that the IVC state is quickly suppressed by the staggered potential and the ground state then becomes the $C_2\mathcal{T}$I.
This is understandable since the stagered potential explicitly breaks the $C_2$ sublattice symmetry, which should favor the $C_2\mathcal{T}$ breaking state.

\subsection{$\nu=-\frac{1}{2}$ case}
At the half filling case, there are two or six electrons occupying the eight
moir\'e bands. Since these two cases are related by particle-hole transformation,
we only consider the case with two electrons, i.e. $\nu$=-1/2.
Similarly, the self-consistent mean-field calculation gives three groups of states
which can be viewed as the flavor-polarized version of those states
found at the charge neutral case.

\begin{figure}[h]
\includegraphics[width=\columnwidth,clip=true]{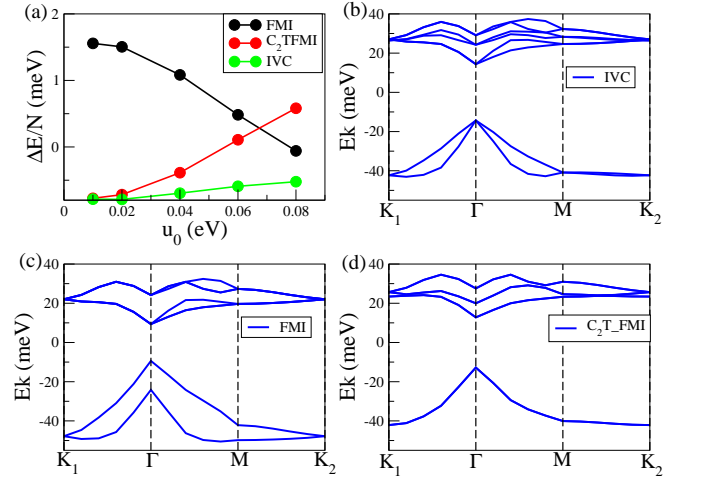}
  \caption{(a) Total energy per particle for ferromagnetic insulator (FMI), flavor polarized $C_2\mathcal{T}$ breaking insulator ($C_2\mathcal{T}$FMI) and intervalley coherent (IVC) states
  at half filling for various values of the intra-sublattice interlayer coupling
  $u_0$, where we choose $\Delta E$ to be relative to the average energy of these three states to clarify the small energy difference. (b)(c)(d) Typical energy dispersion of states of these three groups of states, where $u_0$=0.08 eV and the chemical potential is moved to the center of the gap for convenience.}
  \label{fig:energy_half}
\end{figure}

The first group of states is the fully flavor-polarized states that do not
mix the two moir\'e bands. Since there are only two electrons occupying the eight
bands, only one of the four spin-valley combined flavors is fully filled leaving
the other three completely empty. We again call this group of states as
FMI and the order parameter of FMI can be described by the term
$(s_0 \pm s_z)(\tau_0 \pm \tau_z)\gamma_0$ depending on which combination of flavors
is fully filled.
As shown in panel (c) in Fig.~\ref{fig:energy_half}, the two filled bands below the chemical potential correspond to the two bands of the ordered flavor, and since
$C_2\mathcal{T}$ symmetry is not broken, the Dirac points for each flavor are not destroyed.
One set of the occupied bands is double degenerate corresponding to the spin
degeneracy of the opposite valley of the filled bands.
The energy gap of these states is about 18.7 meV.
The second group of states includes the spin or valley polarized states that also
mixes the two moir\'e bands, so we name them as the $C_2\mathcal{T}$FMI states.
The order parameter of these states can be described by the terms as
$(s_0 \pm s_z)\tau_{0,z}\gamma_x$ for the spin-polarized states
and $s_{0,z}(\tau_0 \pm \tau_z)\gamma_x$ for the valley-polarized states.
The $C_2\mathcal{T}$ symmetry is always broken in the polarized
species (spin or valley) and preserved in the other.
As shown in the panel (d) in Fig.~\ref{fig:energy_half} which corresponds
to the valley-polarized $C_2\mathcal{T}$FMI, the Dirac points
are indeed gapped out in the filled bands, while those in the unoccupied
bands are not destroyed. The energy gap in these states is about 25.5 meV.
The third group of states is the spin-polarized IVC states, which mixes the two
valleys for one spin component. Its order parameter can be described by the
terms as $(s_0 \pm s_z)\tau_{x,y}\gamma_y$. The energy dispersion of this
IVC state is shown in the panel (b) in Fig.~\ref{fig:energy_half},
whose energy gap is about 28.6 meV.

By comparing the total energy per particle of the states in each group, we
again find that the ground state is the IVC state. The energy of the IVC states
is lower by 0.5 meV at $u_0$=0.08 eV and approaches the $C_2\mathcal{T}$FMI as the bandwidth of
the moir\'e bands decreases by tuning down $u_0$.

\begin{figure}[h]
\includegraphics[scale=0.3]{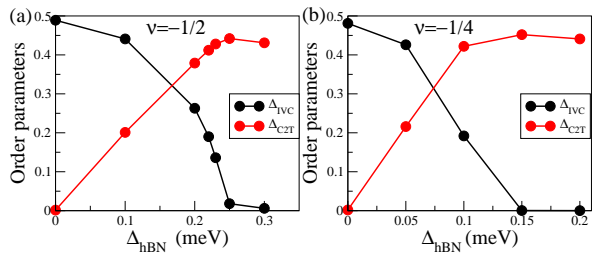}
  \caption{Evolution of the intervalley coherent (IVC) state and $C_2\mathcal{T}$ breaking order parameters with
  the increasing of the staggered potential $\Delta_{hBN}$ for $u_0$=0.08 eV at (a) half filling $\nu=-1/2$  and (b) quarter filling $\nu$=-1/4.}
  \label{fig:ops_delta}
\end{figure}

We have also studied the effect of staggered potential at this filling and we find that by adding the staggered potential $\Delta_{hBN}\sigma_z$,
the IVC ground state starts to mix with the $C_2\mathcal{T}$ breaking order, and the IVC order parameter is quickly suppressed as $\Delta_{hBN}$ reaches 0.3 meV, so that the ground state becomes the $C_2\mathcal{T}$FMI. The evolution of the
two order parameters is shown in the panel (a) in Fig.~\ref{fig:ops_delta}.

\begin{figure}[h]
\includegraphics[width=\columnwidth,clip=true]{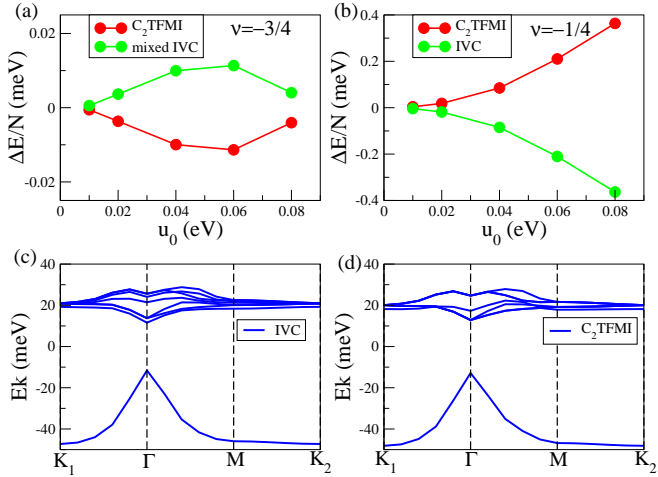}
  \caption{Total energy per particle for flavor-polarized
  C$_2\mathcal{T}$ breaking insulator (C$_2\mathcal{T}$FMI) and mixed intervalley coherent (IVC) states at $\nu=-\frac{3}{4}$ (a) and $\nu=-\frac{1}{4}$ (b) for various values of the intra-sublattice interlayer coupling $u_0$, where we choose $\Delta E$ to be relative to the average energy of these two states to clarify the small energy difference. (c)(d) Typical energy dispersion of states of these two groups of states, where $u_0$=0.08 eV and the chemical potential is moved to the center of the gap for convenience.}
  \label{fig:energy_quarter}
\end{figure}

\subsection{$\nu=-\frac{3}{4}$ case}
At the case $\nu=\pm\frac{3}{4}$, there are one or seven electrons occupying
the eight moir\'e bands. Again, since these two cases are related by particle-hole transformation, we only consider the case with one electron, i.e. $\nu=-\frac{3}{4}$. In this filling, the self-consistent mean-field calculation leads
to two gapped solutions, including the flavor-polarized $C_2\mathcal{T}$FMI
and a mixed order state with both IVC order parameter and $C_2\mathcal{T}$ breaking
order parameter. This time, the energy of these two states are very close to
each other and $C_2\mathcal{T}$FMI is lower in energy by at most 0.022 meV for
various values of $u_0$ from 0.01 eV to 0.08 eV, as
shown in the panel (a) in Fig.~\ref{fig:energy_quarter}.
For the realistic case where $u_0$=0.08 meV, energy gap is about 25.6 meV and for the $C_2\mathcal{T}$FMI state and 23.1 meV for the mixed IVC state as shown in
the panel (c) and (d) in Fig.~\ref{fig:energy_quarter}.
The IVC component of the order parameter of the mixed IVC state will be quickly
suppressed if there is a small staggered potential arising from the alignment
of the hBN substrate and the ground state is $C_2\mathcal{T}$FMI which is the QAH states reported in the experiment~\cite{Serlin_2019,Sharpe_2019}.

\subsection{$\nu=-\frac{1}{4}$ case}
Again we consider only the hole dope side $\nu=-\frac{1}{4}$, i.e. there are three electrons occupying the eight moir\'e bands.
We also find two groups of states in this case, which are the
$C_2\mathcal{T}$FMI state and the IVC state.
In the $C_2\mathcal{T}$FMI state, three out of the four spin-valley flavors
are occupied while the other flavor is completely empty. Within each filled
flavor, the two flat bands are mixed so that $C_2\mathcal{T}$ is breaking
and an energy gap of about 25.7 meV is opened.
The IVC state also only involves three flavors, while two of them with the
same spin and opposite valleys form an IVC state by mixing the two
opposite valleys, the other flavor form a $C_2\mathcal{T}$ breaking order
by mixing the two flat bands within that flavor, so the IVC state at this
filling can also be viewed as the mixed order of the IVC and $C_2\mathcal{T}$
order and the energy gap of this state is about 25.7 meV.
Comparing the energy of the two groups of states, we find that
IVC state is always the ground state and the energy difference from the
$C_2\mathcal{T}$FMI also decreases with the decreasing of the bandwidth
as shown in the panel (b) in Fig.~\ref{fig:energy_quarter}.
Similarly, if we add a small staggered potential $\Delta_{hBN}$, the IVC
order will be suppressed quickly as shown in the panel (b) in Fig.~\ref{fig:ops_delta}.

\section{Comparisons to other works}

In this section, we compare our method and results to the other recent works that also aim at understanding the Hartree-Fock ground states in twisted bilayer graphene
\cite{Liu_2019,Bultinck_2019,Dai_2019,Xie_2020}. Ref. 28 and 29 performed the Hartree-Fock calculations including all remote bands without projecting to the low energy flat bands. This embraces the full complexity of the band structure, and simplifications were made in treating the effects of the Coulomb interaction at the Hartree-Fock level. In our calculation, we project the Coulomb interaction onto the lowest two flat bands per spin and valley, and treat the correlation effect at the Hartree-Fock level more completely. As a consequence, the results we obtained are very different from those of Ref. 28 and Ref. 29. Specifically, Ref. 28 mainly focused on whether the $C_2\mathcal{T}$ symmetry is broken and completely ignored the IVC state, which we find to be the energetically more favorable ground state
at the charge neutral 0, 1/2, and 1/4 fillings. In Ref. 29, the IVC and all order parameters were considered, but an approximation was made to assume that the Fock term is momentum independent. This approximation may lead to different results at the studied 1/2 and 3/4 fillings compared to our results.
For 1/2 filling, we find the ground state is the insulating IVC state consistent with the analytic argument given in Ref. 31, whereas the results in Ref. 29 indicate either an IVC or a valley polarized state or even a coexistence of them as the ground states, which can be either insulating or metallic depending on the screening length and the dielectric constant. For 3/4 filling, our results show an insulating IVC ground state with a finite gap, while the results of Ref. 29 show a metallic state. Since the recent experimental work~\cite{Lu_2019} found that at both 1/2 and 3/4 fillings, the ground state is always an insulating state without hBN alignment, the results obtained here are more in-line with the experiments, suggesting that projection the Coulomb interaction to the lowest flat bands followed by a more complete Hartree-Fock treatment of the correlation effects may capture the low energy physics properly.

The method of projecting the interaction to the low energy flat bands was independently developed in Refs. 30 and 31 to study the charge neutral point at the HF level. Specifically, Ref. 30 projected the Coulomb interaction to the lowest two flat bands, the same as in our approach. However, we found a lower energy ground state, i.e. the IVC state, which was not considered in the calculations of Ref. 30. 

In Ref. 31, the authors projected the Coulomb interaction onto the lowest six bands per spin and valley, and included the IVC state, which was indeed found to be the ground state in their calculations at the charge neutral point. Our results at the charge neutral point is indeed consistent with Ref. 31. For example, the energy gap for the IVC states in our approach is $\Delta_{IVC}=35.5$ meV. The average energies of the IVC and $C_2\mathcal{T}$I states are $E_{IVC}=-21.4$ meV and $E_{C_2\mathcal{T}I}=-20.3$ meV, respectively, implying the IVC state is lower in energy by $1.1$ meV. These results are numerically consistent with those reported in Figs.7(c,d) of Ref. 31, and further indicating that the low energy physics can be captured by projecting the the two lowest flat bands.

The other new result of our work, which is complementary to Ref. 30 and 31, is that we have systematically and quantitatively studied the HF ground states at other commensurate fillings beyond the charge neutron point. This required a generalization of the subtraction scheme to deal with the double counting in the HF calculations introduced in Ref. 30 to ensure the existence of a trivial solution without any symmetry breaking at all commensurate fillings, with the details provided in appendix~\ref{sec:app_hf}. At 1/2 filling, our numerical calculations show that ground state is the spin-polarized IVC state, which is consistent with the analytic argument (explicit calculations were not performed) given in Ref.~\onlinecite{Bultinck_2019}. These agreements at both charge neutral and 1/2 filling not only benchmark our method but also attest the picture that the two emergent flat bands near the magic angle play the most important role, and projecting the Coulomb interaction onto them can determine the ground state properties. The latter may offer a useful framework for implementing more accurate, but numerically costly many-body studies beyond the Hartree-Fock approximation in the future. Moreover, we also obtained the HF ground states at 1/4 and 3/4 fillings, which were not studied in Ref. 31. Finally, the insulating ground states obtained in our calculations at 1/2 and 3/4 fillings are consistent with recent experiments~\cite{Lu_2019} as discussed above.

\section{Conclusion}
\label{sec:conclusion}
In conclusion, we have performed self-consistent mean-field calculations at commensurate filling fractions of the twisted bilayer graphene around the magic angle.
We find insulating ground states for all these fillings, which is
consistent with the experimental results.
For $\nu=0$, $-\frac{1}{2}$ and $-\frac{1}{4}$, the ground state
is the intervalley coherent (IVC) state, and for $-\frac{3}{4}$ filling, the ground state is the flavor-polarized $C_2\mathcal{T}$ breaking insulator ($C_2\mathcal{T}$FMI).
The energy difference between the IVC state and the $C_2\mathcal{T}$I state decreases as the bandwidth of the flat bands decreases.
A small staggered potential associated with the hBN alignment can
quickly suppressed the IVC order and stabilize the $C_2\mathcal{T}$ breaking state for all the fillings.

We have also checked the effect of screening strength by tuning the gate distance $d_s$ in Eq.~\ref{eq:Vg} from 40 nm down to 5 nm. The screening strength is stronger as $d_s$ decreases which makes the Coulomb interaction more short-ranged. We find that as $d_s$ decreases, the total energy of all the mean-field states increases while their energy differences do not change sign, i.e no phase transition occurs. The increase of the energy is due to the fact that the stronger screening reduces the magnitude of the Coulomb interaction therefore reduces the energy gain of the mean-field solution. An example plot for the case with $\nu=-\frac{1}{2}$ is provided in the appendix~\ref{sec:app_ds}.

\textit{Acknowledgments}--
This work is in part supported by the National Natural Science Foundation of China No. 11674278 (YZ and FZ)
and the U.S. Department of Energy, Basic Energy Sciences Grant No. DE-FG02-99ER45747 (KJ and ZW).

\appendix

\begin{widetext}

\section{Projection to the flat bands}
\label{sec:app_proj}
The Bloch wavefunction calculated from the continuum model has the
following form:
\begin{equation}
    \psi_{n,\tau,\bk}^X(\br)=\sum_{\bG}u_{n,\tau;\bG,X }(\bk) e^{i(\bk+\bG)\cdot \br}
    \label{eq:bloch_app}
\end{equation}
where $X=A_1,B_1,A_2,B_2$ corresponds to the mixture of sublattice and layer indices
and $\bG$ is the reciprocal lattice in the morie Brillouin zone expressed
as $\bG=n_1 \bG_1 + n_2 \bG_2$.
Here, we use 121 $\bG$ vectors with $n_1, n_2 \in [-5,5]$ which is enough to produce the flat band.
Eq.~\ref{eq:bloch_app} provides the transformation between the original
monolayer graphene operators $c^{+}$, $c$ and the operators in the projected
eigenbasis $d^{+}$, $d$ which reads:
\begin{equation}
    c_{X,\bk+\bG, s\tau}=\sum_{n} u_{n,\tau;\bG,X }(\bk) d_{n,\bk, s\tau}
\end{equation}
where the momentum of $c$ operator is defined
in the big Brillouin Zone (BZ) of the monolayer graphene while the momentum of
the operator $d$ is defined in the moir\'e Brillouin Zone (mBZ), with the periodic
boundary condition $d_{n,\bk+\bG, s\tau}=d_{n,\bk, s\tau}$
Then the interaction part of the Hamiltonian is
\begin{equation}
\begin{split}
    H_{int}&=\frac{1}{2S}\sum_{X,X', s s',\tau\tau'}\sum_{\tilde{\bk}_1\tilde{\bk}_2\tilde{\bq}\in \text{BZ}} V^{XX'}(\tilde{\bq}) c_{X,\tilde{\bk}_1-\tilde{\bq}  s \tau}^{+}c_{X',\tilde{\bk}_2+\tilde{\bq}  s' \tau'}^+c_{X',\tilde{\bk}_2 s' \tau'} c_{X,\tilde{\bk}_1 s \tau}
    \\
    &=\frac{1}{2S}\sum_{m_1,m_2,m_3,m_4}\sum_{\bk_1\bk_2\bq \in \text{mBZ}}\sum_{\bG_1\bG_2\bG}\sum_{X,X', s s',\tau\tau'} V(\bq+\bG) u^*_{m_1,\tau;\bG_1-\bG,X }(\bk_1-\bq)u^*_{m_2,\tau';\bG_2+\bG,X' }(\bk_2+\bq)
    \\
    &u_{m_3,\tau';\bG_2,X' }(\bk_2)u_{m_4,\tau;\bG_1,X }(\bk_1) d_{m_1\bk_1-\bq  s\tau}^+d_{m_2\bk_2+\bq  s'\tau'}^+d_{m_3\bk_2 s'\tau'}d_{m_4\bk_1 s\tau}
    \\
    &=\frac{1}{2S}\sum_{m_1,m_2,m_3,m_4}\sum_{\bk_1\bk_2\bq \in \text{mBZ}}\sum_{\bG, s s',\tau\tau'} V(\bq+\bG) \lambda_{m_1 m_4; \tau}(\bk_1-\bq,\bk_1+\bG) \lambda^{*}_{m_3 m_2; \tau^{'}}(\bk_2,\bk_2+\bq+\bG)
    \\
     &d_{m_1\bk_1-\bq  s\tau}^+d_{m_2\bk_2+\bq  s'\tau'}^+d_{m_3\bk_2 s'\tau'}d_{m_4\bk_1 s\tau}
    \\
    &=\frac{1}{2S}\sum_{m_1,m_2,m_3,m_4}\sum_{\bk_1\bk_2\bq \in \text{mBZ}}\sum_{ s s',\tau\tau'} V^{\tau\tau'\tau'\tau}_{m_1,m_2,m_3,m_4}(\bk_1,\bk_2,\bq) d_{m_1\bk_1-\bq  s\tau}^+d_{m_2\bk_2+\bq  s'\tau'}^+d_{m_3\bk_2 s'\tau'}d_{m_4\bk_1 s\tau}
\end{split}
\end{equation}
where we use the single-gate-screened Coulomb potential to approximate the
interaction $V^{XX'}(\bq)$ as
\begin{equation}
 V^{XX'}(\bq)=V(\bq)=\frac{e^2}{2\varepsilon\varepsilon_0 q}(1-e^{-2q d_s})
\end{equation}
and $\lambda_{m_1 m_2; \tau}(\bk_1,\bk_2+\bG)$ is the form factor defined by the Bloch wavefunction as
\begin{equation}
 \lambda_{m_1 m_2; \tau}(\bk_1,\bk_2+\bG)=\sum_{\bG^{'},X}u^{*}_{m_1,\tau;\bG^{'},X}(\bk_1)u_{m_2,\tau;\bG+\bG^{'},X}(\bk_2)
\end{equation}
so that the projected Coulomb interaction matrix is
\begin{equation}
 V^{\tau\tau'\tau'\tau}_{m_1,m_2,m_3,m_4}(\bk_1,\bk_2,\bq)= \sum_{\bG}V(\bq+\bG) \lambda_{m_1 m_4; \tau}(\bk_1-\bq,\bk_1+\bG) \lambda^{*}_{m_3 m_2; \tau^{'}}(\bk_2,\bk_2+\bq+\bG)
\end{equation}
The form factor $\lambda_{m_1 m_2; \tau}(\bk_1,\bk_2+\bG)$ satisies the
following two equation as:
\begin{equation}
 \lambda_{m_1 m_2; \tau}(\bk_1,\bk_2+\bG)=\lambda^{*}_{m_2 m_1; \tau}(\bk_2,\bk_1-\bG)
\end{equation}
from the definition, and
\begin{equation}
 \lambda_{m_1 m_2; -\tau}(\bk_1,\bk_2+\bG)=\lambda^{*}_{m_1 m_2; \tau}(-\bk_1,-\bk_2-\bG)
\end{equation}
due to the time reversal symmetry.

\section{Mean-field approximation}
\label{sec:app_hf}
In the section above, we have derived the Eq.~\ref{eq:Hproj}
and ~\ref{eq:hproj} in the main text.
We use the self-consistent HF approximation to decouple the interaction
term. Using the order parameter defined in Eq.~\ref{eq:op}
\[\rho(\bk)_{m_1, s\tau;m_2, s'\tau'}= \left<d_{m_1\bk s\tau}^{+}d_{m_2\bk s'\tau'}\right>\]
where we have confined the order the direction for spin in z-axis, which will
not affect the general conclusion of the ground state, the interaction term
becomes:
\begin{equation}
\begin{split}
    H_{HF}&=\frac{1}{2S}\sum_{\{m_i\}}\sum_{ s s'\tau\tau'}\sum_{\bk_1\bk_2} V_{m_1m_2m_3m_4}^{\tau\tau'\tau'\tau}(\bk_1,\bk_2,0) [\rho(\bk_1)_{m_1, s\tau;m_4, s\tau} d_{m_2\bk_2 s'\tau'}^{+}d_{m_3\bk_2 s'\tau'}
    \\
    &+\rho(\bk_2)_{m_2, s'\tau';m_3, s'\tau'} d_{m_1\bk_1 s\tau}^{+}d_{m_4\bk_1 s\tau} - \rho(\bk_1)_{m_1, s\tau;m_4, s\tau}\rho(\bk_2)_{m_2, s\tau;m_3, s'\tau'}]
    \\
    &-\frac{1}{2S} \sum_{\{m_i\}}\sum_{ s\tau}\sum_{\bk_1\bk_2} V_{m_1m_2m_3m_4}^{\tau\tau'\tau'\tau}(\bk_1,\bk_2,\bk_1-\bk_2) [\rho(\bk_2)_{m_1, s\tau;m_3, s\tau'} d_{m_2\bk_1 s\tau'}^{+}d_{m_4\bk_1 s\tau}
    \\
    &+\rho(\bk_1)_{m_2, s\tau';m_4, s\tau} d_{m_1\bk_2 s\tau}^{+}d_{m_3\bk_2 s\tau'} - \rho(\bk_2)_{m_1, s\tau;m_3, s\tau'} \rho(\bk_1)_{m_2, s\tau';m_4, s\tau}]
    \\
    &=\sum_{\bk s\tau\tau'}\sum_{m_1m_2} h_{m_1,\tau;m_2,\tau',\bk s} d_{m_1\bk s\tau}^{+}d_{m_2\bk s\tau'} - E_c
\end{split}
\end{equation}
where,
\begin{equation}
\begin{split}
    h_{m_1,\tau;m_2,\tau',\bk s}&=\frac{1}{2S}\sum_{m_1'm_2'\bk'} \{ \sum_{ s'\tau''}[V_{m_1'm_1m_2m_2'}^{\tau''\tau\tau\tau''}(\bk',\bk,0)+V_{m_1m_1'm_2'm_2}^{\tau\tau''\tau''\tau}(\bk,\bk',0)] \rho(\bk')_{m_1', s'\tau'';m_2', s'\tau''}
    \\
    &-[V_{m_1'm_1m_2'm_2}^{\tau'\tau\tau\tau'}(\bk,\bk',\bk-\bk')+V_{m_1m_1'm_2m_2'}^{\tau\tau'\tau'\tau}(\bk',\bk,\bk'-\bk)] \rho(\bk')_{m_1', s\tau';m_2', s\tau} \}
    \\
    &=\frac{1}{S}\sum_{m_1'm_2'\bk'} \{ \sum_{ s'\tau''}V_{m_1m_1'm_2'm_2}^{\tau\tau''\tau''\tau}(\bk,\bk',0) \rho(\bk')_{m_1', s'\tau'';m_2', s'\tau''} \delta_{\tau\tau'}
    -V_{m_1'm_1m_2'm_2}^{\tau'\tau\tau\tau'}(\bk,\bk',\bk-\bk') \rho(\bk')_{m_1', s\tau';m_2', s\tau} \}
\end{split}
\label{eq:hhf}
\end{equation}
and
\begin{equation}
\begin{split}
    E_c&=\frac{1}{2S}\sum_{\{m_i\}}\sum_{\bk_1\bk_2, s\tau\tau'} [\sum_{ s'} V_{m_1m_2m_3m_4}^{\tau\tau'\tau'\tau}(\bk_1,\bk_2,0) \rho_{m_1,m_4,\bk_1 s\tau} \rho_{m_2,m_3,\bk_2 s'\tau'}
    \\
    &- V_{m_1m_2m_3m_4}^{\tau\tau'\tau'\tau}(\bk_1,\bk_2,\bk_1-\bk_2) \rho(\bk_2)_{m_1, s\tau;m_3, s\tau'} \rho(\bk_1)_{m_2, s\tau';m_4, s\tau}]
\end{split}
\end{equation}
is the condensation energy.
The first term in Eq.~\ref{eq:hhf} is the Hartree term and the second term
is the Fock term.
Then the mean-field Hamiltonian becomes:
\begin{equation}
 H_{MF}=\sum_{\bk s\tau\tau'}\sum_{m_1m_2} [(\epsilon_{m_1\bk\tau}-\mu)\delta_{m_1m_2}\delta_{\tau\tau'}+ h_{m_1,\tau;m_2,\tau',\bk s}] d_{m_1\bk s\tau}^{+}d_{m_2\bk s\tau'} - E_c
 \label{eq:hmf}
\end{equation}
For each momentum $\bk$ and spin $ s$, $h_{m_1,\tau;m_2,\tau',\bk s}$
is a 4 $\times$ 4 matrix. In the self-consistent HF calculation, we start with some initial values of the density matrix $\rho(\bk)$, and then solve the
Eq.~\ref{eq:hmf} for the new $\rho(\bk)$, which is used to calculate the
new $H_{MF}$ until $\rho(\bk)$ is converged.

Since the density matrix $\rho(\bk)$ is finite even for the no-symmetry breaking states at the charge channel, which will be double-counted when it couples to the density operators in the mean-field Hamiltonian.
In order to address this issue, we have used a scheme where we replace
the density matrix $\rho(\bk)$ by $\tilde{\rho}(\bk)=\rho(\bk)-\rho_0(\bk)$
in Eq.~\ref{eq:hhf}
where, $\rho_0(\bk)$ is the density matrix of the no-symmetry breaking
state at filling studied. Using this scheme, the mean-field Hamiltonian
will always have a trivial solution that corresponds to the no-symmetry
breaking state at any particular filling that is studied. At the charge
neutrality, our approach coincides with the one used in Ref.~\onlinecite{Liu_2019}. Note, Ref.~\onlinecite{Bultinck_2019} also studied the charge neutral point numerically, where they used a slightly different $\rho_0(\bk)$, which corresponds to the density matrix for the two decoupled graphene layers at the charge neutral point.

\section{Energy difference as a function of gate distance $d_s$}
\label{sec:app_ds}
We have also checked the effect of screening strength by tuning the gate distance $d_s$ in Eq.~\ref{eq:Vg} from 40 nm down to 5 nm. We observe similar behavior for all the communsurate fillings.  As shown in Fig.~\ref{fig:e_d},  the total energy of all the mean-field states increases as $d_s$ descreases, while their energy differences do not change sign, i.e the IVC state is always the ground state as $d_s$ decreases from 40 nm to 5nm.

\begin{figure}[h]
\includegraphics[scale=0.4]{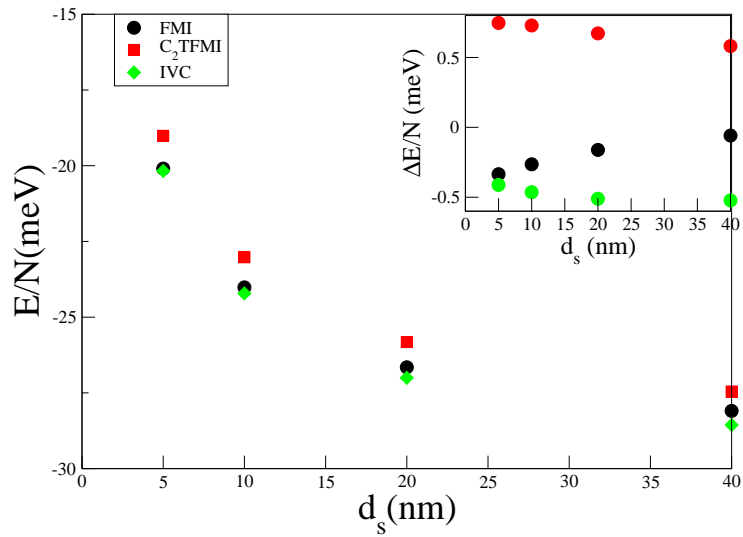}
  \caption{Main plot: Total energy per particle for ferromagnetic insulator (FMI), flavor polarized $C_2\mathcal{T}$ breaking insulator ($C_2\mathcal{T}$FMI) and intervalley coherent (IVC) states
  at half filling for various values of the gate distance
  $d_s$. Inset: the relative energy $\Delta E$ to the average energy of these three states to clarify the small energy difference.}
  \label{fig:e_d}
\end{figure}

\end{widetext}

\bibliography{reference}
\end{document}